\documentclass[twocolumn,showpacs,preprintnumbers,amsmath,amssymb]{revtex4}

\usepackage{graphicx}
\usepackage{dcolumn}
\usepackage{bm}

\begin{document}

\preprint{Ver21}

\title{Ferromagnetic coupling and magnetic anisotropy in molecular Ni(II)
squares}

\author{R. Koch}
\author{O. Waldmann}
\author{P. M\"uller}
\affiliation{ Physikalisches Institut III, Universit\"at
Erlangen-N\"urnberg, D-91058 Erlangen, Germany }
\author{U. Reimann}
\author{R. W. Saalfrank}
\affiliation{ Institut f\"ur Organische Chemie, Universit\"at
Erlangen-N\"urnberg, D-91054 Erlangen, Germany }

\date{\today}

\begin{abstract}
We investigated the magnetic properties of two isostructural
Ni(II) metal complexes [Ni$_{4}$L$^b_{8}$] and
[Ni$_{4}$L$^c_{8}$]. In each molecule the four Ni(II) centers form
almost perfect regular squares. Magnetic coupling and anisotropy
of single crystals were examined by magnetization measurements and
in particular by high-field torque magnetometry at low
temperatures. The data were analyzed in terms of an effective spin
Hamiltonian appropriate for Ni(II) centers. For both compounds, we
found a weak intramolecular ferromagnetic coupling of the four
Ni(II) spins and sizable single-ion anisotropies of the easy-axis
type. The coupling strengths are roughly identical for both
compounds, whereas the zero-field-splitting parameters are
significantly different. Possible reasons for this observation are
discussed.
\end{abstract}


\maketitle

\section{\label{sec:introduction}Introduction}

Modern inorganic chemistry provides arrangements of magnetic metal
ions in highly symmetric geometries, where the metal centers are
separated by organic ligands. Compounds with a topologically
simple arrangement of their metal centers like grids and rings
have been investigated extensively.
\cite{Co22g,Cu33g,Mn33g,Fe10Lip,Fe6Gatt,CsFe8} Due to their high
symmetry it is possible to experimentally determine the magnetic
parameters of large and rather complex systems. As intermolecular
effects are negligible in these systems, they actually form
perfect models to explore finite size spin systems. \cite{Ni22g}
In a number of cases, several related species of a compound class
are available, allowing to observe correlations between
crystallographic structure and magnetic properties.
\cite{BarraFe8,CsFe8} Investigation of such correlations is of key
help for understanding e.g. coupling mechanisms in detail.
\cite{Kahn2}

Recently, the highly symmetric Ni(II) square [Ni$_{4}$L$^b_{8}$]
with L$^b$ = C$_5$H$_4$N-CON-CN$_4$-C$_2$H$_5$ attracted
considerable interest. \cite{Ni22s,NibNic} Preliminary magnetic
studies of powder samples revealed a sizeable ferromagnetic
coupling of the four Ni(II) metal centers within a molecule. This
system is thus one of the very rare examples of a ferromagnetic
Ni(II) complex. A second species, [Ni$_{4}$L$^c_{8}$] with L$^c$ =
C$_5$H$_4$N-CSN-CN$_4$-C$_2$H$_5$, could be also synthesized.
\cite{URDiss,NibNic} The two ligands used differ only at one
position, i.e the oxygen in L$^b$ is replaced by a sulfur in
L$^c$. Interestingly, the two Ni$_4$ squares are not only
isostructural. Their magnetically relevant geometrical dimensions
actually differ by less than 3.5\%. This leads to a new situation:
Differences in magnetic properties stem predominately from
different electronic properties of oxygen and sulfur and not from
geometrical distinctions. It is the purpose of this work to
investigate the magnetism, especially the magnetic coupling and
anisotropy, of these two compounds in detail.

When focusing on anisotropic properties, several experimental
methods like SQUID magnetometry, EPR and torque magnetometry are
possible. Torque magnetometry has been proven to be a very
valuable tool in this field. \cite{Fe6Gatt,CsFe8,Cu33g}

\begin{figure*}
\includegraphics{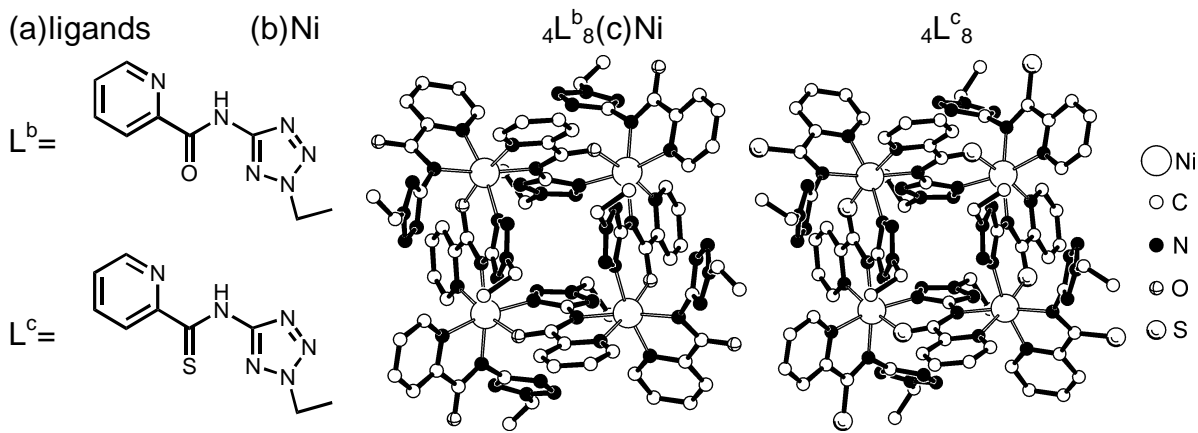}
\caption{\label{fig:kristallstruktur} (a) Sketch of both ligands L$^b$ and
L$^c$. (b) Structural representation of [Ni$_{4}$L$^b_{8}$] and (c)
[Ni$_{4}$L$^c_{8}$] (view along the crystallographic $S_4$-axis, H atoms
omitted).}
\end{figure*}

\begin{table}
\caption{\label{tab:diffs}Comparison of selected distances and bonding
angles of the coordination sphere for the compounds [Ni$_{4}$L$^b_{8}$] and
[Ni$_{4}$L$^c_{8}$]. Considering the nickel centers at the upper right
corners in Figs.~\ref{fig:kristallstruktur}(b) and
\ref{fig:kristallstruktur}(c), respectively, the N atoms of their
coordination spheres were numbered clockwise from N1 to N5 starting with
oxygen/sulfur.}
\begin{ruledtabular}
\begin{tabular}{lcr}
 & [Ni$_{4}$L$^b_{8}$] & [Ni$_{4}$L$^c_{8}$]\\ \hline Ni - O/S & 2.032(2)
\r{A} & 2.051(3) \r{A} \\ Ni -
N1 & 2.081(3) \r{A} & 2.073(4) \r{A} \\ Ni - N2 & 2.076(3) \r{A} & 2.072(4)
\r{A} \\ Ni - N3
 & 2.111(3) \r{A} & 2.074(4) \r{A} \\  Ni - N4 & 2.065(3) \r{A} & 2.112(4)
\r{A} \\ Ni - N5 &
2.062(3) \r{A} & 2.070(4) \r{A}\\ &  & \\  O/S - Ni - N1 &
88.44(11) $^\circ$ & 86.98(15) $^\circ$ \\ N1 - Ni - N2 &
79.02(13) $^\circ$ & 78.04(17) $^\circ$ \\ N2 - Ni - N3 &
94.62(12) $^\circ$ & 98.04(12) $^\circ$ \\ N3 - Ni - N4 &
79.26(11) $^\circ$ & 79.20(15) $^\circ$ \\ N4 - Ni - N5 &
90.28(12) $^\circ$ & 92.46(15) $^\circ$ \\ N5 - Ni - O/S &
83.24(11) $^\circ$ & 83.30(14) $^\circ$ \\
\end{tabular}
\end{ruledtabular}
\end{table}

\section{\label{sec:experimental}Experimental}
\subsection{\label{sec:preparation}Preparation and Crystal Structures}

The tetranuclear Ni(II) cluster [Ni$_{4}$L$^b_{8}$] $\cdot$
4CH$_2$Cl$_2$ with L$^b$ = C$_5$H$_4$N-CON-CN$_4$-C$_2$H$_5$ was
prepared as described in Ref.~\onlinecite{Ni22s}. The
isostructural compound [Ni$_{4}$L$^c_{8}$] $\cdot$ 4CH$_2$Cl$_2$
with L$^c$ = C$_5$H$_4$N-CSN-CN$_4$-C$_2$H$_5$ was synthesized
with a method analogous to that used for the [Ni$_{4}$L$^b_{8}$]
complex. \cite{URDiss} The two ligands L$^b$ and L$^c$ are
sketched in Fig.~\ref{fig:kristallstruktur}. They only differ by
one position: The C=O group in L$^b$ is replaced by a C=S group in
L$^c$.

The crystal structures were determined by X-ray structure analysis
of single crystals. \cite{Ni22s,URDiss} Both compounds crystallize
in the space group $I$4(1)/a. They exhibit crystallographic $S_4$
molecular symmetry with the four nickel centers forming almost
regular squares (Fig.~\ref{fig:kristallstruktur}). The molecular
$S_4$ symmetry axes are perpendicular to the planes of the
molecules defined by the Ni centers, and necessarily coincide with
the magnetic $z$-axes of the molecules. The shape of the crystals
is quadratic bipyramidal. The molecular magnetic $z$-axes are thus
parallel to the $S_4$ symmetry axis of the crystals.

The eight ligands in a molecule coordinate the nickel centers in
two different ways (Fig.~\ref{fig:kristallstruktur}). A set of
four ligands links two nickel centers each and builds up the
[Ni$_4$L$_4$]$^{4+}$ square-like cores. The second set of four
ligands coordinates the nickel centers at the corners of the
square cores. Each nickel center is surrounded by six donor atoms,
five N and one O for [Ni$_{4}$L$^b_8$] and five N and one S for
[Ni$_{4}$L$^c_8$], respectively, forming slightly distorted
octahedral coordination spheres.

Although the sulfur donors are significantly larger than the
oxygen donors, the structures of the two complexes are remarkably
similar. This is evident from a careful inspection of
Figs.~\ref{fig:kristallstruktur}(b) and
\ref{fig:kristallstruktur}(c). The most notable structural
difference arises for the CN$_4$-C$_2$H$_5$ groups of the
corner-ligands. Their orientations differ slightly in the two
compounds. This is plausible since these groups are not
coordinated to nickel centers.

However, the geometry of the nickel coordination sphere as well as
the structure of the ligand linking two nickel ions are almost not
affected by the replacement of oxygen with sulfur (Table
\ref{tab:diffs}). The Ni-Ni next neighbor distance is
5.567(5)$\,$\r{A} for [Ni$_{4}$L$^b_8$] and 5.560(5)$\,$\r{A} for
[Ni$_{4}$L$^c_8$]. Several further distances and bond angles are
listed in Table~\ref{tab:diffs} for both compounds.

These structural elements are most relevant for the magnetic
properties, i.e. ligand-field and superexchange interactions. As
the two molecules are almost structurally identical, the
potentially different magnetic properties of [Ni$_{4}$L$^b_8$] and
[Ni$_{4}$L$^c_8$] should be controlled predominantly by the
different electronic properties of oxygen and sulfur.

\subsection{\label{sec:magnmeas}Magnetization Measurements}

For magnetization measurements, a single crystal was selected by
light microscopy in the mother liquor. To avoid decomposition, the
crystal was transferred directly from the solution into Apiezon
grease and mounted on a plastic straw. The weight of the crystals
was typically 100$\,\mu$g. The background signal of grease and
sample holder was negligible compared to the signal of the
crystals. Magnetic moment was measured with a MPMS-7 SQUID
magnetometer from Quantum Design. The temperature range was 1.8-
300$\,$K, the maximum magnetic field 5.5$\,$T. The measurements
were performed for magnetic fields parallel and perpendicular to
the $z$-axis of the crystal. Due to the quadratic bipyramidal
shape of the single crystals, none of the crystal planes is
parallel to the magnetic $z$-axis. Therefore, a proper orientation
of the sample on the sample holder was difficult. The orientation
accuracy was about 10$^{\circ}$. Two samples of each compound were
investigated.

\subsection{\label{sec:extorquemagnetometry}Magnetic Torque Measurements}

\begin{figure}
\includegraphics{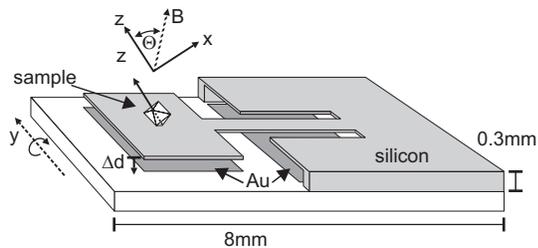}
\caption{\label{fig:torqueschema} Sketch of the silicon
torquemeter described in section~\ref{sec:sensor}. The typical
orientation of a quadratic bipyramidially shaped crystal sample
and its $z$-axis are shown. For adjusting the angle $\Theta$, the
device could be rotated in-situ around the magnetic $y$-axis.}
\end{figure}

Torque measurements of single crystal samples were performed with
an appropriately designed silicon cantilever torque sensor which
will be described in detail in the next chapter. As for
magnetization measurements, a crystal was selected from the
mother liquor, immediately covered with grease, and mounted on the
torque sensor. The typical sample weight was less than 10$\,\mu$g,
but could be determined only within an error of 50\% due to the
unknown amount of grease covering the crystals. Therefore, the
number of molecules in a crystal, $X$, is known only roughly and a
calibration of the torque signal was not feasible. The torquemeter
was mounted on a rotating plate which allowed an in-situ
orientation of the crystal with an accuracy of 0.3$^{\circ}$. The
sample was mounted on the sensor with its $z$-axis perpendicular
to the rotation axis, as indicated in Fig.~\ref{fig:torqueschema}.
The positioning accuracy was better than 3$^\circ$, which is much
better than for the magnetization measurements.
Figure~\ref{fig:torqueschema} defines the angle $\Theta$ between
magnetic field and magnetic $z$-axis of the crystal. The
torquemeter provides a resolution of 10$^{-11}\,Nm$. It was
inserted into a 15/17$\,$T superconducting cryomagnet system. The
temperature was adjusted by a variable temperature insert.
Typically, torque measurements versus applied magnetic field were
performed at 15 different angles $\Theta$ in a range of
220$^{\circ}$. The lowest temperature was 1.8$\,$K. For several
samples, the low-field range was investigated in detail, i.e. at
60 different angles in a range of $180^{\circ}$. 5 samples of each
compound were investigated.

\section{\label{sec:sensor}Silicon Cantilever Torquemeter}

A schematic drawing of the cantilever device is presented in
Fig.~\ref{fig:torqueschema}. The cantilever is mounted on a
substrate of crystalline quartz glass. It is made from a very pure
crystalline silicon wafer (resistivity $>3\,k\Omega cm$) with a
(100) surface orientation and thickness of 300\,$\mu$m. The use of
crystalline silicon guarantees excellent mechanical as well as
non-magnetical properties. As shown in
Fig.~\ref{fig:torqueschema}, the single crystal silicon cantilever
consists of parts at the original thickness of 300\,$\mu$m and
strongly thinned parts. This has been achieved by different masks
on the two sides of the wafer. The masks were formed by 1\,$\mu$m
thick $SiO_2$ layers grown on both sides of the silicon wafer
which were patterned by standard photolithography and etching with
buffered fluorine acid. The silicon cantilever itself was etched
by hot KOH to a thickness of 10-30\,$\mu$m. A 200\,nm gold layer
was evaporated on the bottom side of the cantilever. Together with
gold pads structured on the quartz substrate it forms a capacitor
and a reference capacitor (which is placed at the right hand
section of the device in Fig.~\ref{fig:torqueschema}). Finally,
the cantilever was glued on the quartz substrate.

The torque of the sample causes a deflection of the cantilever
which is detected by a change of the capacitance $\Delta C$. For
readout, the two capacitors were connected to a ratio transformer
forming an ac bridge. With this setup a sensitivity of $\Delta
C/C_0=10^{-7}$ is readily obtained. \cite{Rich1,Wald1} $C_0$
denotes the zero-field capacitance (1\,pF). The properties of the
cantilever torquemeter can be modeled as follows: The deflection
is $\Delta d=(3/2)\tau/(D_c L_c)$, where $D_c$ is the spring
constant and $L_c$ the length of the cantilever. The change of the
capacitance is given by $\Delta C/C_0 \approx \Delta
d/d_0(1+\Delta d/d_0)$, where $d_0$ denotes the distance of the
capacitor plates. This relation shows nonlinear behavior. However,
for all measurements presented in this work nonlinearity was less
than 2\% and could be neglected. With $U_0$ being a characteristic
of the ac bridge, its output voltage can be expressed as
$U=U_0(\Delta C/C_0)$. Altogether, one obtains $U=K\tau (1+\alpha
K\tau)$ with the calibration constant K and the nonlinearity
$\alpha$. If required, K and $\alpha$ may be obtained from an
explicit calibration which can be done quite easily in many ways.
\cite{Wald1,Schwarz1,XFe6} As we were not able to determine the
weight of the samples accurately, calibration of the device was
not necessary.

\section{\label{sec:theory}Theory}

The appropriate Hamiltonian for molecular spin clusters consisting
of Ni(II) centers is given by \cite{Ben90,Ni22g}

\begin{eqnarray}
H=-\sum_{i<j}J_{ij}\mathbf S_i\cdot \mathbf S_j+\cr\sum_i\mathbf
S_i\cdot \mathbf D_{i}^{lig}\cdot \mathbf S_i+ \mu _B
\sum_i\mathbf S_i\cdot \mathbf g_i\cdot \mathbf B
\end{eqnarray}

with $\mathbf S_i = 1$ and the following standard terms: a
Heisenberg term modeling isotropic next-neighbor exchange
interactions, a zero-field- splitting (ZFS) term due to
ligand-field interactions, and the Zeeman term. Due to the small
coupling constants in [Ni$_{4}$L$^b_{8}$] and [Ni$_{4}$L$^c_{8}$],
anisotropic and biquadratic coupling terms can be neglected.
Dipole-dipole interactions are negligible due to the large
distances of the spin centers. Cross-coupling superexchange terms
hardly exist as corresponding coupling paths are not present. Due
to the molecular $S_4$ symmetry of the complexes, magnetic
anisotropy is strictly uniaxial and a simplified Hamiltonian is
obtained:

\begin{eqnarray}
\label{eqn:hamiltonian} H=-J\sum_{i<j}\mathbf S_i\cdot \mathbf
S_j+ D\sum_i( S_{i,z}^2-2/3)+\cr \mu _Bg_{xy}(S_xB_x+S_yB_y)+\mu
_Bg_zS_zB_z.
\end{eqnarray}

Thus, four magnetic parameters are sufficient to describe the
properties of the Ni$_4$ squares correctly: the coupling constant
$J$, the ZFS parameter $D$ and the two $g$-factors $g_{xy}$ and
$g_z$. In the following we will use the parameterization $g =
\sqrt{(2 g_{xy}^2 + g_z^2)/3}$ and $\Delta g = g_z-g_{xy}$ for the
$g$-factors. As we will see, $J \approx 0.9\,$K and $D \approx
-2.5\,$K. Thus neither the strong exchange limit $(|D/J|\ll 1)$
nor the Ising limit $(|D/J|\gg 1)$ is valid. Therefore, a full
matrix diagonalization has to be performed. Due to the small
dimension of the Hilbert space of $3^4=81$, this can be done on a
commercial PC. Calculation time could be reduced by a factor of 12
by taking into account a $C_{2\nu}$ spin permutational symmetry of
Hamiltonian eq.~(\ref{eqn:hamiltonian}). \cite{Symmetrie}

\begin{figure}
\includegraphics{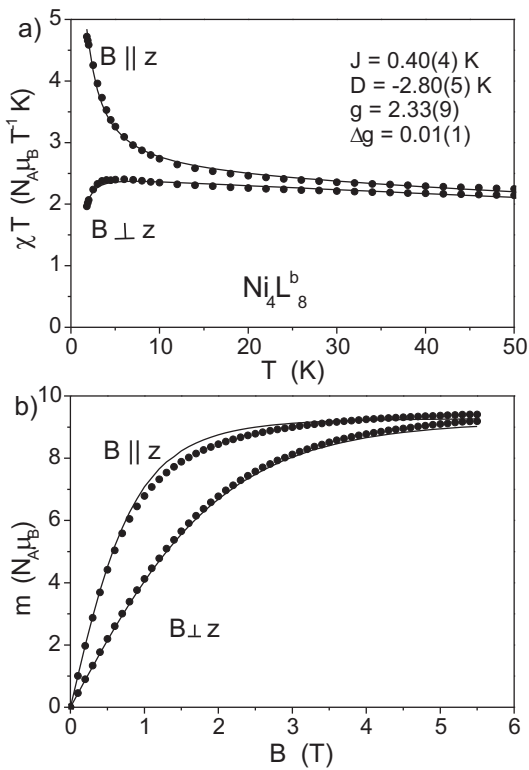}
\caption{\label{fig:squid} (a) Magnetic susceptibility times
temperature vs. temperature and (b) magnetic moment vs. magnetic
field at T\,=\,1.8\,K of a [Ni$_{4}$L$^b_{8}$] crystal for two
different orientations of magnetic field. The solid lines
represent best fits based on eq.~(\ref{eqn:hamiltonian}).}
\end{figure}

\section{\label{sec:magnmeasresults}Magnetization Measurements: Results and
Analysis}

Magnetic susceptibility and magnetization curves for
[Ni$_4$L$^b_8$] with magnetic field parallel and perpendicular to
the magnetic $z$-axis are shown in Fig.~\ref{fig:squid}. Fits
based on eq.~(\ref{eqn:hamiltonian}) revealed the following
parameters: $J = 0.40(4)\,$K, $D=-2.80(5)\,$K, $g=2.33(9)$ and
$\Delta g=0.01(1)$ for [Ni$_4$L$^b_8$] and $J=0.40(4)\,$K,
$D=-1.80(8)\,$K, $g=2.33(9)$ and $\Delta g= 0.01(1)$ for
[Ni$_4$L$^c_8$]. It should be noted, that the given errors reflect
statistical uncertainties only and do not include systematical
errors. Detailed simulations showed that the low temperature
behavior of the magnetization is very sensitive to a misalignment
of the crystal. Already a misalignment of $5^{\circ}$, which is
well within experimental uncertainty (see
section~\ref{sec:magnmeas}), leads to notably different
parameters. Therefore, the results from the magnetization
measurements will be regarded as guidelines and will not be
discussed further. Nevertheless, a trend is indicated: The
coupling constants are roughly identical in both compounds,
whereas the ZFS parameters differ significantly.

\begin{figure}
\includegraphics{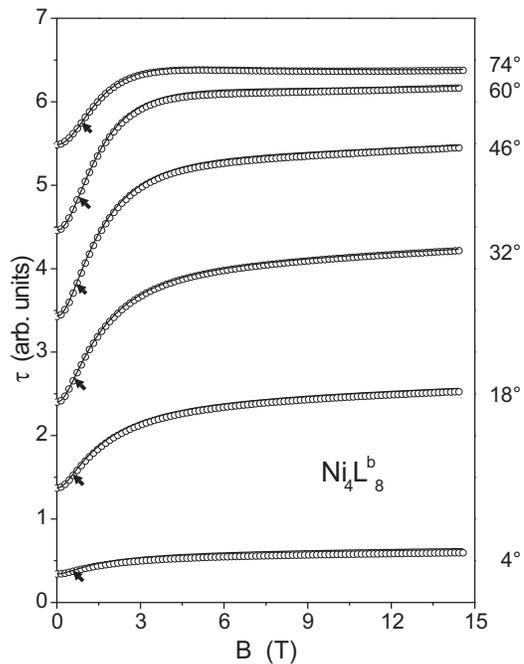}
\caption{\label{fig:torquemessung} Torque measurements vs.
magnetic field of a [Ni$_{4}$L$^b_{8}$] crystal sample at several
angles $\Theta$ and $T$ = 1.8\,K (circles). The curves are shifted
for clarity. The solid lines represent fits using Hamiltonian
eq.~(\ref{eqn:hamiltonian}). The arrows indicate the inflection
points of the curves.}
\end{figure}

\section{\label{sec:torquemagnetometry}Torque Magnetometry: Results and
Analysis}

The torque measurements were analyzed by fitting Hamiltonian
eq.~(\ref{eqn:hamiltonian}) to the data. As the torque signal was
uncalibrated, the number $X$ of molecules in a sample had to be
considered also as a free parameter, in addition to the parameters
$J$, $D$, $g$, and $\Delta g$. In our first attempt to determine
the magnetic parameters, we fitted the complete angular dependence
$\tau$(B, $\Theta$) of a sample with $X$, $J$, $D$ and $\Delta g$
varying simultaneously. $g$ was fixed to reasonable values between
2.1 and 2.3. \cite{Abr70,Ni_g_fak,Ni22g} A typical data set with
corresponding fit is shown in Fig.~\ref{fig:torquemessung}.
Concerning the parameter R=$\sum{(\tau_{sim}-\tau_{meas})^2}/
\sum{\tau_{meas}^2}$, which estimates the quality of the least
square fits, we obtained excellent results. Unfortunately, this
approach revealed rather large variations of the parameters for
different samples. However, fixing one additional parameter
besides $g$ yielded stable fitting results. Thus, either $J$, $D$
or $\Delta g$ should be determined in a different way. As we will
show, this can be done without additional experimental data. The
improved strategy rests on the fact, that a fit to the whole data
set does not consider that the parameters affect individual parts
of the torque curves selectively.  A careful study of numerical
simulations based on Hamiltonian eq.~(\ref{eqn:hamiltonian})
provides valuable information concerning this topic.

Figure~\ref{fig:squid}(b) shows that the magnetization curves
saturate at fields of about 5\,T. This is easily understood as
calculations show that the ground state is well separated from the
exited states at 5\,T ($\Delta$E$>$5\,K). The torque measurements
exhibit similar saturation (Fig.~\ref{fig:torquemessung}). In this
field regime, the torque curves are featureless and the magnitude
is controlled by $X$, $\Delta g$, and $D$ simultaneously. Here the
curves are over parameterized.

\begin{figure}
\includegraphics{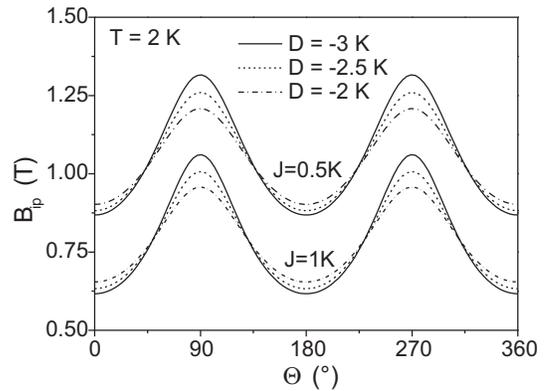}
\caption{\label{fig:winkelsimu} Calculated values for the
inflection point B$_{ip}$ of torque curves as function of $\Theta$
for different parameters $J$ and $D$ ($g = 2.2$, $\Delta g = 0$
).}
\end{figure}

For a detailed examination of the low-field part, it is useful to
analyze the behavior of the inflection points B$_{ip}$($\Theta$)
of the torque curves. In this way, the parameter $X$ is eliminated
as the value of B$_{ip}$($\Theta$) is independent of the magnitude
of the curves. Figure~\ref{fig:torquemessung} shows that the
inflection points vary slightly with $\Theta$. The calculated
angular dependence of B$_{ip}$ for different values of $J$ and $D$
is presented in Fig.~\ref{fig:winkelsimu}. It turns out that the
offset of the oscillation is shifted downwards with increasing $J$
(and $g$, not shown here). On the other hand, the offset is not
influenced by $D$ and, as further simulations showed, by $\Delta
g$. In contrast, the amplitude of the oscillation is controlled by
$D$ exclusively. So it is possible to extract the values of $D$
and $J$ (for given $g$) from analyzing the inflection points of
the measured torque curves. As $g$ should be between 2.1 and 2.3,
$J$ may be obtained with an accuracy of 10\%.

\begin{figure}
\includegraphics{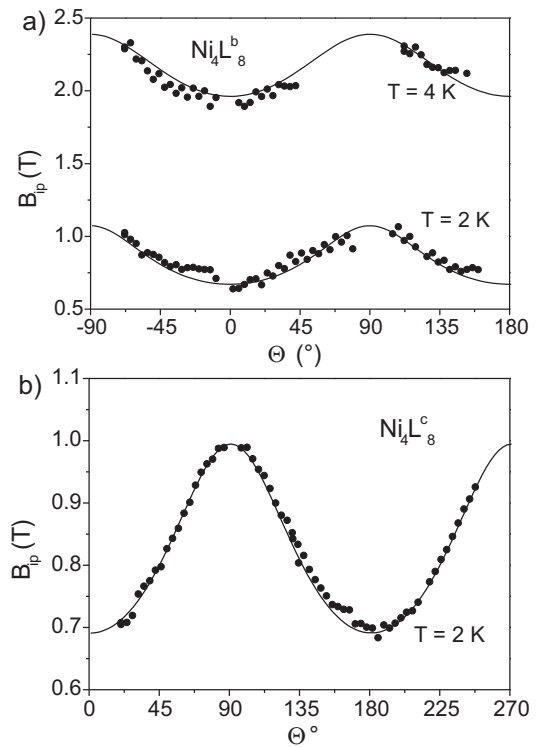}
\caption{\label{fig:winkelabh} B$_{ip}$ vs. angle $\Theta$ of
measured torque curves for (a) [Ni$_{4}$L$^b_{8}$] and (b)
[Ni$_{4}$L$^c_{8}$]. Solid lines represent fits using
eq.~(\ref{eqn:hamiltonian}). For [Ni$_{4}$L$^b_{8}$], data sets
for two temperatures have been used simultaneously in the
analysis.}
\end{figure}

For two samples of every compound, we determined
B$_{ip}$($\Theta$) from measurements at 60 different angles.
Fitting this data (Fig.~\ref{fig:winkelabh}) revealed $D$ and $J$
very precisely. To obtain the remaining parameter $\Delta g$, the
whole torque data set $\tau$(B, $\Theta$) was fitted with $D$ and
$g$ fixed. As $J$ was not fixed in this procedure, it could be
determined again and compared to the value extracted from
B$_{ip}$($\Theta$). This provided prove for the consistency of our
analysis. The parameters for the two compounds are summarized in
Table~\ref{tab:results}.

\begin{table}
\caption{\label{tab:results}Magnetic parameters for the examined
molecules.}
\begin{ruledtabular}
\begin{tabular}{lcccr}
&J (K)&D (K)&$\Delta$ g\\ \hline
[Ni$_{4}$L$^b_{8}$]&0.9(1)&-2.7(1)&0.01(1)\\
$[$Ni$_{4}$L$^c_{8}]$&0.9(1)&-2.0(1)&0.01(1)\\
\end{tabular}
\end{ruledtabular}
\end{table}

\begin{figure}
\includegraphics{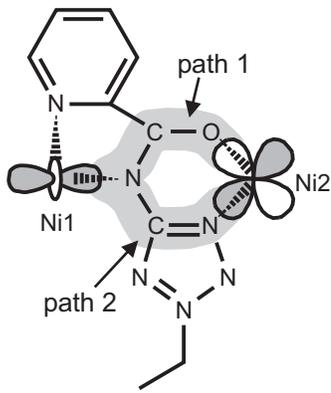}
\caption{\label{fig:orbitals} Sketch of the ligand L$^b_{8}$
bridging two nickel ions, the free metal orbitals of the two
linked Ni centers, and the two obvious coupling paths.}
\end{figure}

\section{\label{sec:discussion}Discussion}

The value of $J$ observed for powder samples of [Ni$_4$L$^b_8$] in
Ref.~\onlinecite{Ni22s} is almost a factor of two larger than that
determined here. We ascribe this to structural changes of the
cluster resulting from loss of CH$_2$Cl$_2$ molecules upon drying
of crystals. This effect has been observed for other molecules,
which decompose rapidly under exposure to air, too.
\cite{XFe6,CsFe8} The different values for $J$ obtained from
single crystal SQUID measurements as compared to the torque
results is explained by misalignment errors as discussed in
section~\ref{sec:magnmeasresults}.

Within the error ranges, the value of $\Delta g$ is consistent
with predictions of ligand-field theory: From $\Delta g =
2D/\lambda$ and $\lambda=-250\,$K appropriate for Ni(II) centers
\cite{Abr70} one obtains $\Delta g$=0.02.

The above analysis demonstrated that the coupling constants are
identical within experimental accuracy in both compounds, whereas
the ZFS parameters differ significantly. The geometry of the Ni
coordination spheres are essentially identical for the two
complexes. Thus, different values of $D$ should be ascribed to
different electronic environments of the Ni centers. In
particular, the different donor capabilities of oxygen and sulfur
should clearly affect the ligand-fields and thus the ZFS
parameters.

The results for $J$ are more puzzling. The special geometrical
arrangement of the two coordination pockets leads to an
othogonality of the metal orbitals. To point this out,
Fig.~\ref{fig:orbitals} shows a sketch of the ligand L$^b$ linking
two Ni centers and the relevant metal orbitals (hypothetical
orbitals of a free Ni atom). The magnetic orbitals should extend
along the coupling path between neighboring Ni ions as indicated
by the gray background in Fig.~\ref{fig:orbitals}. The coupling
path actually may be split in two sub-paths, path 1 along the
Ni-N-C-O-Ni chain and path 2 along the Ni-N-C-N-Ni chain. This
suggests an explanation for the observed ferromagnetic couplings:
When path 1 and path 2 contribute equally to the magnetic
coupling, the overlap of the magnetic orbitals will be zero for
parity reasons. Then, magnetic coupling would be ferromagnetic
since an antiferromagnetic contribution, which is proportional to
the overlap, cancels out. \cite{Kahn1}

However, path 1 of the coupling path is quite different for
[Ni$_{4}$L$^b_{8}$] and [Ni$_{4}$L$^c_{8}$], as demonstrated by
the observed different values for $D$. Thus, one would expect that
also the coupling constants are clearly different, in contrast to
experimental observation. This suggests that the magnetic coupling
is predominantly controlled by path 2. But then, an
antiferromagnetic coupling should be expected which generally
dominates if the overlap is non-zero. \cite{Kahn1} Obviously, the
ferromagnetic coupling in [Ni$_{4}$L$^b_{8}$] and
[Ni$_{4}$L$^c_{8}$] is the result of a subtle balance between
ferromagnetic and antiferromagnetic contributions which is not
easily reconciled with present knowledge.

\section{\label{sec:conclusion}Conclusion}

The two Ni$_4$ compounds studied in this work are of interest for
three reasons: (i) Ferromagnetic coupling in molecular spin
systems is rather rare. Concerning polynuclear Ni complexes, only
few systems are known to date. \cite{Ni4Kers} (ii) The high
crystallographic symmetry reduces the number of magnetic
parameters significantly, allowing very accurate determination of
the magnetic parameters. (iii) Two species with minimal
geometrical differences of one system facilitates an isolation of
possible links between ligand-field splitting or magnetic
coupling, respectively, and electro-structural properties.

We showed that $J$ and $D$ can be determined very accurately by
torque magnetometry in combination with sophisticated analysis
also for systems where $J$ and $D$ are on the same order of
magnitude. This extends recent applications of torque magnetometry
to molecular nanomagnets. \cite{Fe6Gatt,CsFe8,Cu33g}

The differences of the anisotropy parameter $D$ have been ascribed
to local differences in the electronic environment of the Ni
centers. For the coupling constant $J$ some mechanisms have been
suggested but no final explanation could be given. Ab-initio
calculations would be of great help to provide comprehensive
explanations for the origin and the strength of the coupling.
\cite{KorV15}

\begin{acknowledgments}
We would like to thank Andreas Richter for help in the lab and
Jochen Thomas for silicon handling. Furthermore we thank Stefan
Schromm and Stephan Rother for valuable discussions. This work was
supported by the Deutsche Forschungsgemeinschaft.
\end{acknowledgments}

\bibliography{my}

\end{document}